\documentclass[a4paper,aps,pra,%twocolumn,
preprintnumbers]{revtex4}
\usepackage{graphics,graphicx,epsfig}

\usepackage[centertags]{amsmath}
\usepackage{amsfonts}
\usepackage{amssymb}
\usepackage[centertags]{amsmath}
\usepackage{amsfonts}
\usepackage{amssymb}
\usepackage{amsthm}
\newcommand{\beq}{\begin{equation}}
\newcommand{\eeq}{\end{equation}}
\newcommand{\beqa}{\begin{eqnarray}}
\newcommand{\eeqa}{\end{eqnarray}}

\newcommand{\ket} [1] {\vert #1 \rangle}
\newcommand{\bra} [1] {\langle #1 \vert}

\newcommand{\proj}[1]{\ket{#1}\bra{#1}}
\newcommand{\mean}[1]{\langle #1 \rangle}

\begin{document}

%\date{}
\sloppy
\begin{abstract}
Multiparticle entangled states generated via interaction between
narrow-band light and an ensemble of identical two-level atoms are
considered. Depending on the initial photon statistics, correlation
between atoms and photons can give rise to entangled states of these
systems. It is found that the state of any pair of atoms interacting
with weak single-mode squeezed light is inseparable and robust
against decay. Optical schemes for preparing entangled states of
atomic ensembles by projective measurement are described.
\end{abstract}
%\pacs{PACS numbers: 03.67.Mn}
\title{On Multiparticle Entanglement via Resonant Interaction
between Light and Atomic Ensembles}
\author{V.N. Gorbachev, A.I. Trubilko}.
\affiliation {Laboratory for quantum information  $\&$
computation,
University of AeroSpase Instrumentation,\\
St-Petersburg, 190000, Bolshaia Morskaia 67, Russia.}

%\begin{document}
\maketitle
%\tableofcontents
\section{Introduction}
 Now properties of multiparticle entangled states, their
preparation and application are the subject of extensive discussion.
The desired state of a physical system can be prepared either by
projective measurement or as a result of evolution. For atomic
systems, both methods have already been implemented in experiments.
In particular, two atomic ensembles were used in \cite{Pol} to
create an EPR pair by projective measurement. The latter method was
demonstrated in several studies: entangled states of alkali ions
were generated via Coulomb interaction \cite{M}, neutral Rydberg
atom were used to create an EPR pair in a micromaser setup
\cite{HMN}, and resonant dipole - dipole interaction was used for
entangling neutral atoms in an optical lattice \cite{Lat}. The most
popular methods for preparing entangled photon states are still
mostly based on parametric down-conversion. For example, an
entangled state equivalent to a three-state quantum system (qutrit)
was prepared and examined by using quantum state tomography in
\cite{MS}. These examples suggest that an entangled state of two
systems can be prepared experimentally by using a certain
interaction. Systems of this kind are well studied. With regard to
applications, it is important to know how entanglement can be
utilized and its robustness against decoherence. In this respect, of
special interest are multiparticle systems, whose entangled states
are characterized by much more complicated and diverse behavior.

Previous efforts were mainly focused on analysis of entanglement
between several particles. In particular, the W class of tripartite
entanglement defined in \cite{6} includes the symmetric three-photon
polarization entangled state implemented in the experiment reported
in \cite{7}. An extension to four qubits was proposed in \cite{8},
where nine inequivalent classes were distinguished that cannot be
connected by local operations and quantum communication. Studies of
multiparticle systems are relatively few, being focused on
entanglement criteria and application to problems in quantum
information theory. Whereas the Peres - Horodecki criterion for
bipartite entanglement found in \cite{9} was applied to a real
physical system in \cite{10}, no operational criterion is known for
entanglement in the general case, and various approaches are often
used. In \cite{11}, the concept of entanglement molecules \cite{12}
was introduced to propose a classification using graphs, in which
particles and classical or quantum correlations represented,
respectively, by vertices and edges connecting pairs of vertices.
Graphs of this kind can be used to describe both pure and mixed
entangled states and distinguish several classes differing by
topological properties of the graphs. In \cite{13}, symmetric states
(including Dicke states) were studied by using several entanglement
measures (entanglement entropy, negativity, and entanglement of
formation) defined by the eigenvalues of a partial transpose of the
density matrix. A numerical analysis was performed to find that
symmetric states are robust to particle loss even if the number of
particles is large (up to $10^{3}$ ). Note that the calculation of
eigenvalues is a difficult task, because the dimension of an
ensemble’s Hilbert space exponentially increases with the number of
constituent particles. Owing to their robustness, symmetric states
can be used in such applications as cloning and telecloning
protocols for quantum information transmission \cite{14}, quantum
key distribution \cite{15}, and quantum teleportation or dense
coding \cite{16}. The formulation of two models of a one-way quantum
computer using measurements on multiparticle entangled states
\cite{17}, \cite{18} has strongly stimulated studies of the
properties of multiparticle systems, in particular, Ising- and Bose
- Hubbard-like models.

The present study focuses on the Dicke states arising as a result of
collective interaction of many atoms with electromagnetic field
\cite{19}, which has been analyzed in numerous studies (e.g., see
\cite{20}). This system exhibits many physical properties of
interest for quantum information processing. Photon trapping in
chain configurations of atoms was considered in \cite{21}. When the
system is placed in a cavity, this effect reduces the photon escape
rate and increases the decoherence time of the cavity mode. In
\cite{22}, this effect was used for generating W states and
anticloning \cite{23}, which can be implemented with high fidelity
by means of photon trapping. In those studies, only single-photon
traps and single-photon initial states were analyzed. Here, we
consider the more general case of multiphoton processes, assuming
that the photon statistics is arbitrary.

The main questions addressed below are the following: what types of
entangled states are produced by interaction between atoms and
field? What states can be prepared from independent atomic ensembles
entangled with a photon? How can these states be utilized? We
consider resonant interaction between narrow-band light and an
ensemble of identical two-level atoms coupled to a common heat bath.
The analysis is restricted to a simple model of radiative decay.
Multiphoton processes, such as Raman scattering, are described in
terms of effective Hamiltonians, which can be obtained by unitary
transformation \cite{24}. The behavior of an atomic system
interacting with light characterized by arbitrary photon statistics
is analyzed by using perturbation theory in the interaction strength
for arbitrary statistics oh light particularly for Gaussian,
coherent, and squeezed states. We find that weak single- mode
squeezed light is required to create multiparticle entanglement
between atoms. As distinct to the case considered in \cite{25}, the
steady state discussed here is robust against atomic decay. When
decay is neglected and analysis is restricted to a single-photon
initial state, simple exact solutions describing exchange of
excitation between the field mode and atoms can be obtained
\cite{26}. These solutions can be used for generating and
transforming symmetric Dicke states and for processing and storing
quantum information. The optical schemes for projective measurement
considered here can be used to generate entangled states of atomic
ensembles. An EPR entangled pair of macroscopic ensembles was
created in an experiment \cite{Pol}. The new states produced in our
schemes have hierarchical structure, thus differing from the cluster
states introduced in \cite{27} as a resource for one-way computing.

The paper is organized as follows. First, we formulate a basic model
and write out the second-order perturbation solutions obtained by
taking into account radiative decay. These solutions are then used
to analyze the states of the atomic system corresponding to various
photon statistics. Exact solutions obtained under certain initial
conditions by neglecting radiative decay are used to describe
generation and transformation of symmetric Dicke states. Finally, we
consider optical schemes for preparing entangled states of atomic
ensembles by projective measurement.

\section{Basic equations}
In the dipole approximation, the ensemble of $N$ identical, but
distinguishable, two-level atoms interacting with electromagnetic
field is described by the Hamiltonian
\begin{eqnarray}
\nonumber
H=i\hbar\vartheta, &&\\
\nonumber \vartheta=\sum_{k}
%\epsilon_{m}
g_{k}a_{k}S^{\dagger}_{k}-h.c.,&&
\end{eqnarray}
where
$g_{k}=(\hbar\omega_{k}/2\varepsilon_{0}L^{3})^{1/2}(\mu,e_{k})$ is
the coupling constant, $\mu$ is the dipole transition matrix
element, $e_{k}$ is the polarization vector for the mode with wave
vector $k$, $a_{k}$ and $a_{k}^{\dagger}$  are photon creation and
annihilation operators,
$S^{\dagger}_{k}=\sum_{a}s_{10}(a)\exp(ikr_{a})$ is the atomic
operator for the atom located at a point $r_{a}$ (x,y = 0, 1, where
0 and 1 denote the ground and excited states, respectively). When
analysis is restricted to interaction with a single resonant mode,
$S_{k}$ can be replaced with $S_{k=0}$, which makes it possible to
treat an atomic ensemble occupying a spatial region as a point like
object. Then
\begin{eqnarray}
\label{001}
 \vartheta=S_{10}B-S_{01}B^{\dagger},
\end{eqnarray}
where $S_{10}=\sum_{a}\ket{1}_{a}\bra{0}$, $B=ga$. Effective
Hamiltonian (\ref{001}) is used here to describe not only
interaction with a single resonant mode, but also multiphoton
processes, such as Raman scattering. In the latter case, we set
$B=fa_{A}a^{\dagger}_{S}$, and assume that the photon frequencies
$\omega_{A}$ and $\omega_{S}$ satisfy the relation
$\omega=\omega_{A}-\omega_{S}$, where $\omega$ is the atomic
transition frequency. Hamiltonians of this form can be obtained by
unitary transformation \cite{24}.
\\
The density matrix $\rho$ of  $N$ atoms interacting with field obeys
the master equation
\begin{eqnarray}
\label{002} \frac {\partial} {\partial t}
\rho=[\vartheta,\rho]+\mathcal{L}\rho,
\end{eqnarray}
where relaxation is represented by the Lindblad superoperator
\begin{eqnarray}
\nonumber
 \mathcal{L}=\sum_{a}\mathcal{L}_{a},
\end{eqnarray}
\begin{eqnarray}
\mathcal{L}_{a}=-\frac{\gamma_{\uparrow}}{2}[s_{01}(a)s_{10}(a)\rho
-s_{10}(a)\rho
s_{01}(a)]-\frac{\gamma_{\downarrow}}{2}[s_{10}(a)s_{01}(a)\rho-
s_{01}(a)\rho s_{10}(a)]+h.c.&&
\end{eqnarray}
This representation corresponds to the model of purely radiative
decay with longitudinal and transverse decay rates
$\gamma=\gamma_{\downarrow}+\gamma_{\uparrow}$ and $\gamma_{\bot}$ ,
which satisfy the relation $\gamma_{\bot}=\gamma/2$. In general case
$\gamma_{\bot}>\gamma/2$ since $\gamma_{\bot}$ should be replaced by
$\gamma_{\bot}+\kappa$, where  $\kappa$ is a dephasing collision
rate.

Effective Hamiltonian (\ref{001}) may involve many field modes with
 $\omega_{k}$ differing from the atomic transition frequency
 by $\delta \omega_{k}$ and
occupying a frequency band of width $\Delta \omega$. If $\Delta
\omega, \delta \omega_{k}\ll \gamma_{\bot}$, then we can consider a
narrow- band radiation field and make use of resonance
approximation. Otherwise, the field must be described in terms of
multiple-time correlation functions. Solution of Eq. (\ref{002}) is
a difficult task. To describe the interaction between single atom
and field, the following equation for the density matrix
$\rho_{a}=Tr'_{a}\rho$ is derived from (\ref{002}) by tracing  over
all atoms except for one:
\begin{eqnarray}
 \label{003}
\frac {\partial} {\partial t}
\rho_{a}=[\vartheta_{a},\rho_{a}]+\mathcal{L}_{a}\rho_{a}+
N(N-1)Sp_{a'}[\vartheta_{a'},\rho_{aa'}],
\end{eqnarray}
where $\vartheta_{a}=s_{10}(a)B-h.c.$ and $\rho_{aa'}=Sp'_{aa'}\rho$
is a two-particle density matrix. The right-hand side of (\ref{003})
contains a multiparticle contribution proportional to $N(N-1)$,
because the density matrix $\rho_{aa'}$  does not commute with the
field operators. This leads to the Bogolyubov-
Born-Green-Kirkwood-Yvon chain of equations for the multiparticle
density matrices $\rho_{a},\rho_{aa'},\rho_{aa'a''},\dots$. In
physical terms, this means that fluctuations of quantized
electromagnetic field induce correlation between atoms. If the field
is assumed to be classical and noise-free, for example, a coherent
state is considered, then the interaction will not give rise to any
correlation, and the initially uncorrelated atoms will remain
mutually independent. In what follows, we use (\ref{002}) to analyze
interactions that can be used to generate symmetric Dicke states.

\section{Dicke states}

First, we define symmetric Dicke states and introduce a
representation of symmetric Dicke states that demonstrates their
relation to the collective interaction processes. The Dicke states
are eigenstates of the operators $J_{z}$ and
$J^{2}=J_{x}^{2}+J_{y}^{2}+J_{z}^{2}$
\begin{eqnarray}
\nonumber
J_{z}\ket{jma}=m\ket{jma}&&\\
J^{2}\ket{jma}=j(j+1)\ket{jma},&&
\end{eqnarray}
where $[J_{s},J_{p}]=i\epsilon_{spd}J_{d}$. For example, operators
$J_{s}, s=x,y,z$ can be represented by Pauli matrices
$J_{s}=(1/2)\sum_{k}\sigma_{sk}$, $\sigma_{sk}, s=x,y,z$. Indexes
$j$ and $m$ are integer or half-integer numbers $|m|\leq j, \max
j=N/2$. If $j=N/2$, then the states are symmetric, and the quantum
number “a” introduced to lift degeneracy can be omitted. For h
excited atoms $h=m+N/2$, the states can be represented as
\begin{eqnarray} \label{0021}
\ket{j=N/2,m} \equiv \ket{h;N}=
\sum_{z}P_{z}\ket{1_{1},1_{2},\dots,1_{h},0_{h+1},\dots,0},&&
\end{eqnarray}
where  $P_{z}$ is one of the $C_{h}^{N}=N!/(h!(N-h)!)$
distinguishable permutations of particles.

The vector $\ket{h;N}$ describes an atomic of $h$ excited atoms and
it is normalized as $\mean{h;N|h;N}=C_{h}^{N}$. Symmetric states of
a multiparticle system arise when interaction is described by
collective operators of the form
$S_{10}=\sum_{a}^{N}\ket{1}_{a}\bra{0}$.
\begin{equation}
\label{007} \ket{h;N}=(1/h!)S_{10}^{h}\ket{0;N}.
\end{equation}
If $h=1$, then one finds that
\begin{eqnarray}
\ket{1;N} =\ket{10\dots0}+\dots +\ket{00\dots1}.
\end{eqnarray}
Since the wavefunction $\ket{h; N}$ is not factorizable, it
represents an entangled state. In terms of correlation between
particles, it is substantially different from other entangled
states. For example, in the Greenberger– Horne–Zeilinger (GHZ) state
$GHZ=(1/2)(\ket{0}^{\otimes N}+\ket{1}^{\otimes N})$, the
correlation of any $M$ particles $(M < N)$ is classical. In
particular, the density matrix corresponding to the state $\proj{1;
N}$ of a group of $M$ particles is $\rho(M\leq
N)=N^{-1}\proj{1;M}+(N-M)N^{-1}\proj{0;N}$. The corresponding von
Neumann entropy depends on the relative particle number $p = M/N$:
$S(\rho(M\leq N))=-p\log p-(1-p)\log (1-p)$. When $p=1/2$ the
entropy achieves its maximum 1. If $M=2$ we can apply the necessary
and sufficient separability criterion proposed in \cite{9}.
According to this criterion, the state is inseparable (entangled) if
the density matrix partially transposed over the one of the atoms
has at least one negative eigenvalue. In the case considered here,
one of the four eigenvalues $\{1/N; 1/N; (N-2)(2N)^{-1}[1 \pm
\sqrt{1+4/(N-2)^{2}}]\}$ is negative. Note that the behavior of
correlation between $M$ particles depends on $p = M/N$. As the total
particle number $N$ increases, $p\to 0$ and the correlation
vanishes, since their state becomes pure as $\rho(M\leq
N)\to\proj{0;N}$ In what follows, we make use of the following
equalities:
\begin{eqnarray}
\label{DD} \nonumber
S_{01}\ket{0;N}=0,&&\\
\nonumber
 S_{10}\ket{h;N}=(h+1)\ket{h+1;N},&&\\
 \nonumber
 S_{01}\ket{h;N}=(N-h+1)\ket{h-1;N},&&\\
 \nonumber
 S_{01}S_{10}\ket{h;N}=(h+1)(N-h)\ket{h;N},&&\\
 S_{10}S_{01}\ket{h;N}=h(N-h+1)\ket{h;N}.&&
\end{eqnarray}
\section{Second-order perturbation theory}

To solve Eq. (\ref{002}), we use perturbation theory in the
interaction strength:
\begin{eqnarray}
\rho=\rho^{(0)}+\rho^{(1)}+\rho^{(2)}+\dots,
\end{eqnarray}
Here, the zeroth-order approximation $\rho^{(0)}$ is the
steady-state solution of (\ref{002}) with $\vartheta=0$:
$\rho^{(0)}=\proj{0}\otimes\rho_{f}$, where the density matrix
$\rho_{f}$ represents the modes and $\ket{0}=\ket{0}^{\otimes N}$
corresponds to the ground state of all atoms. The operators
$\rho^{(k)}$, $k=1,\dots$ satisfy the equations
\begin{eqnarray}
\frac {\partial} {\partial t}
\rho^{(k)}=[\vartheta,\rho^{(k-1)}]+\mathcal{L}\rho^{(k)},
\end{eqnarray}
subject to the initial conditions  $\rho^{(k)}(0)=0$.

The analysis that follows is restricted to second-order perturbation
theory, which is sufficient to obtain statistical characteristics of
the excitation field. The matrix equation for $\rho^{(2)}$ is
\begin{eqnarray}
\nonumber \bra{1_{k},1_{m};N} \frac {\partial} {\partial t}
\rho^{(2)}\ket{0;N}=
-2\gamma_{\bot}\bra{1_{k},1_{m};N}\rho^{(2)}\ket{0}+
\bra{1_{k},1_{m};N}R\ket{0}&&\\
\nonumber \bra{1_{k};N} \frac {\partial} {\partial t}
\rho^{(2)}\ket{1_{m};N}=
-2\gamma_{\bot}\bra{1_{k};N}\rho^{(2)}\ket{1_{m};N}+
\bra{1_{k};N}R\ket{1_{m};N}, ~~~k\neq m&&\\
\nonumber \bra{1_{k};N}\frac {\partial} {\partial t}
\rho^{(2)}\ket{1_{k};N}=
-\gamma\bra{1_{k};N}\rho^{(2)}\ket{1_{k};N}+
\bra{1_{k};N}R\ket{1_{k};N}&&\\
\bra{0;N}\frac {\partial} {\partial t} \rho^{(2)}\ket{0;N}=
\gamma\sum_{k}\bra{1_{k};N}\rho^{(2)}\ket{1_{k};N}+
\bra{0;N}R\ket{0;N}.&&
\end{eqnarray}
where $s_{10}(k)\ket{0;N}=\ket{1_{k};N}$ and
$s_{10}(k)s_{10}(p)\ket{0;N}=\ket{1_{k},1_{p};N}$ represent the
states in which only the kth atom is excited and only the kth and
pth atoms are excited, respectively. The nonzero matrix elements of
the operator $R=[\vartheta,\rho^{(1)}]$ are
\begin{eqnarray}
\nonumber
\bra{1_{k},1_{m};N}R\ket{0;N}=2\kappa(t)B^{2}\rho_{f},&&\\
\nonumber
\bra{0;N}R\ket{0;N}=-\kappa(t)N(B^{\dagger}B\rho_{f}+\rho_{f}B^{\dagger}B),&&\\
\bra{1_{k};N}R\ket{1_{m};N}=2\kappa(t)B\rho_{sf}B^{\dagger},&&
\end{eqnarray}
where $\kappa(t)=(1/\gamma_{\bot})(1-\exp(-\gamma_{\bot}t)).$ For
purely radiative decay, $\gamma_{\bot}=\gamma/2$ and the
second-order perturbation theory yields
\begin{eqnarray}\label{0025}
\nonumber \rho=\proj{0}\otimes
\rho_{f}+\kappa[\ket{1;N}\bra{0;N}\otimes
B\rho_{sf}+h.c.]+\kappa^{2}[\ket{2;N}\bra{0;N}\otimes
B^{2}\rho_{f}+h.c.]
&&\\
\nonumber -N\gamma
\mathcal{K}\proj{0;N}\otimes[B^{\dagger}B\rho_{f}-
B\rho_{f}B^{\dagger}+h.c.]-(1/2)N\kappa^{2}\proj{0;N}\otimes
[B^{\dagger}B\rho_{f}+h.c.]&&\\
+\kappa^{2}\proj{1;N}\otimes B\rho_{f}B^{\dagger},&&
\end{eqnarray}
where
$$ \mathcal{K}=  \gamma_{\bot}^{-1} \left\{\gamma^{-2}[\gamma t+1-\exp(-\gamma t)]
-\frac{\kappa^{2}}{2} \right\}.$$ This expression is valid to second
order if the field is relatively weak:
\begin{equation}\label{00250}
N\kappa^{2}\mean{B^{\dagger}B}\ll 1.
\end{equation}
In the case of interaction with a single resonant cavity mode, we
have $B=ga$ and $\kappa^{2}\mean{B^{\dagger}B}=n/n_{s}$, where
$n_{s}=(\gamma_{\bot}/g)^{2}$ is a saturation parameter and
$n=\mean{a^{\dagger}a}$ is the mean photon number. Then,
(\ref{00250}) reduces to the standard condition imposed in the case
of resonant coupling between the field and two-level atoms: $
Nn/n_{s}\ll 1$. Solution (\ref{0025}) describes the joint evolution
of the atomic ensemble and field starting from an ensemble of
ground-state atoms and an arbitrary state of the field.

\section{Mixed nonseparable atomic states}

Second-order perturbation theory predicts correlation between atoms
depending on photon statistics, i.e., provides a framework for
describing entangled (inseparable) atomic states. To analyze the
properties of the atomic system, we use second-order perturbation
theory to find the density matrix for a group of $M\leq N$ atoms,
$\rho_{A}(M\leq N)$, obtained by taking the trace of (\ref{0025})
over the field states represented by $\rho_{f}$ and over $N-M$
particles. The result has the form
\begin{eqnarray}
\label{320A} \nonumber \rho_{A}(M\leq N)
 \nonumber
=\ket{0}\bra{0}[1-M\kappa^{2}\mean{B^{\dagger}B}]
+\kappa[\mean{B}\ket{1;M}\bra{0}+h.c.]
 \nonumber
+\kappa^{2}[
\mean{B^{2}}\ket{2;M}\bra{0}+h.c.]&&\\
+\kappa^{2}\mean{B^{\dagger}B} \ket{1;M}\bra{1;M}.&&
\end{eqnarray}
Note that the density matrix $\rho_{A}(M\leq N)$ describes a mixed
state of the atomic ensemble. Unlike the density matrices for
symmetric Dicke states (\ref{0021}), $\rho_{A}(M\leq N)$ is
independent of both $N$ and $p = M/N$. Therefore, the correlations
between $M < N$ atoms are identical and are independent of the total
particle number. This implies that the state is robust to particle
loss.

The atomic density matrix cannot be factorized because of the
correlation depending on photon statistics. Consider two atoms
described in terms of their respective observables $c_1$ and $c_2$
such that $[c_{1}; c_{2}] = 0$. Setting $M = 2$ in (\ref{320A}), we
have the two-atom density matrix
\begin{eqnarray}\label{322}
\nonumber
\rho_{A}(2)=\proj{00}(1-2\kappa^{2}\mean{B^{\dagger}B})+\kappa\mean{B}
\Big(\ket{10}\bra{00}+\ket{01}\bra{00}+h.c.\Big)+\kappa^{2}\mean{B^{2}}
\Big(\ket{11}\bra{00}+h.c.\Big)&&\\
%\nonumber
+\kappa^{2}\mean{B^{\dagger}B}\Big(\ket{10}+\ket{01}\Big)\Big(
\bra{10}+\bra{01} \Big).&&
\end{eqnarray}
Using (\ref{322}) we find that the covariance of the operators
$c_{1},c_{2}$ is determined by the electromagnetic field variance:
\begin{eqnarray} \label{Cov}
%\nonumber
\mean{c_{1}c_{2}}-\mean{c_{1}}\mean{c_{2}}=\kappa^{2}[(\mean{B^{2}}-\mean{B}^{2})
\bra{0}c_{1}\ket{1}\bra{0}c_{2}\ket{1}%&&\\
%\nonumber
+(\mean{B^{\dagger}B} -\mean{B^{\dagger}}\mean{B})
\bra{1}c_{1}\ket{0}
\bra{0}c_{2}\ket{1}%&&\\
%\nonumber
+c.c].&&
 \end{eqnarray}
If the field is not fluctuating in the sense that its variances are
zero, i.e., $\mean{B^{2}}-\mean{B}^{2}=0$ etc. (which is true in the
present case, e.g., for a coherent state), then there is no
correlation between atoms. Suppose that $c_{k} (k = 1, 2)$ are
dipole operators: $c_{k}=d_{k}=\mu(s_{01}(k)+s_{10}(k))$, where the
matrix element $\mu$ is real. Then the correlation between two
dipole moments depends on photon statistics. We define the
quadrature operator $X_{f}=B^{\dagger}\exp(i\theta)+h.c.$. Then
(\ref{Cov}) implies that the covariance of the dipole moments is
determined by the variance of the quadrature operator normally
ordered with respect to the field operators $B$ and $B^{\dagger}$ at
$\theta= 0$: $ \mean{d_{1}d_{2}}-\mean{d_{1}}\mean{d_{2}}=
\mu^{2}\kappa^{2}D_{N} $, where
$D_{N}=\mean{X_{f}^{2}}-\mean{X_{f}}^{2}- \mean{[B,B^{\dagger}]}$.
For coherent states, the variance is $D_{N}=0$. The dipole moments
are correlated both for a squeezed-state field (with $D_{N}<0$) and
for field in a classical state (with $D_{N}>0$).

The necessary and sufficient condition for inseparability of a mixed
state is provided by the Peres – Horodecki criterion \cite{9}, which
is valid for systems with Hilbert spaces of dimension $2\times 2$
and $2\times 3$. In the case considered here, the state of a
two-atom system described by $\rho_{A}(2)$ is inseparable
(entangled) if at least one eigenvalue of the density matrix
partially transposed over the variable of atom 1
$\rho_{A}(2)^{T_{1}}$ is negative. As example, we consider light in
Gaussian and squeezed states.

For a Gaussian field ($\mean{B}=\mean{B^{2}}=0$, expression
(\ref{322}) reduces to the density matrix describing a superposition
of the ground and mixed states:
$\rho_{A}(2)=a\proj{00}+b[(\ket{01}+\ket{10})(\bra{01}+\bra{10})]$,
where $a+2b=1$ and $a=1-2\kappa^{2}\mean{B^{\dagger}B}$. The
eigenvalues of  $\rho_{A}(2)^{T_{1}}$ are
\begin{eqnarray*}
 \lambda=\Big\{b,b,
 \frac{a}{2}\pm\sqrt{\frac{a^{2}}{4}+b^{2}}\Big\}.&&
\end{eqnarray*}
Since $\sqrt{a^{2}/4+b^{2}}\approx a/2$, in the approximation
considered here, we have the eigenvalues: $\{b,b,a,0\}$ i.e., a
separable state.

Consider the case of resonant interaction with single- mode squeezed
light $(B = ga)$ generated, for example, by a parametric oscillator.
A simple model of the oscillator is defined by the effective
Hamiltonian $H=i\hbar(f/2)(a^{\dagger 2}-h.c.)$. The solution is
$a=a_{0}\cosh r+a_{0}^{\dagger}\sinh r$, where  $r=f\tau$ is the
squeezing parameter, $\tau$ is the normalized length of the
nonlinear medium, and $a_{0}, a_{0}^{\dagger}$ and denote the input
field operators. For the initial vacuum state, $\mean{a}=0$,
$\mean{a^{2}}=\mean{a^{\dagger 2}}=\cosh r\sinh r$,
$\mean{a^{\dagger}a}=\sinh^{2} r$. In this case (\ref{322}) reduces
to the following the two-atom density matrix
\begin{eqnarray}\label{334}
\nonumber \rho_{A}(2)=\proj{00}[1-2\kappa^{2}\mean{B^{\dagger}B}] +
\kappa^{2}\Big[\mean{B^{2}}
\Big(\ket{11}\bra{00}+\ket{00}\bra{11}\Big)+h.c.\Big]&&\\
+\kappa^{2}\mean{B^{\dagger}B}\Big(\ket{10}\bra{10}+\ket{01}\bra{10}+
\ket{10}\bra{01}+\ket{01}\bra{01}\Big).&&
\end{eqnarray}
The four eigenvalues of $\rho_{A}^{T_{1}}(2)$ are
\begin{equation}\label{336}
\lambda=\left\{0;~~ 1-\frac{2}{n_{s}}\sinh^{2}r;~~ \pm
\frac{1}{n_{s}}\exp(\pm r)\sinh r\right\}.
\end{equation}
To be specific, we set $r > 0$, i.e., consider the state squeezed
with respect to canonical momentum or phase. In this case,
$(-1/n_{s})\sinh r\exp(-2r)<0$. However, it is clear that the degree
of squeezing is low, because the approximations used here imply that
\begin{equation}\label{337}
\frac{\sinh^{2}r}{n_{s}}\ll 1.
\end{equation}
Thus, the state of the atomic system is inseparable. This behavior
is explained as follows. Fluctuations of light give rise to
correlation between atoms, which leads to two-atom coherence. When
condition (\ref{337}) holds, this coherence plays the key role.
Since absorption is weak, the system is almost entirely in the
ground state. As distinct to the case of Gaussian statistics, the
density matrix has the form
$\rho_{A}(2)\approx\ket{00}\bra{00}+\kappa^{2}
[\mean{B^{2}}\ket{11}\bra{00}+h.c.].$

Note that the following two observations can be inferred from this
example. First, a steady entangled atomic state can be created by
using weak squeezed light, which looks promising from an
experimental perspective. Second, the entire ensemble cannot be
interpreted as separable, because any pair in a group of $M\leq N$
atoms is entangled, i.e., the quantum correlation of the ensemble as
a whole is robust to particle loss

Since no reliable universally applicable criterion is known for
multiparticle entanglement, we apply the Peres – Horodecki criterion
to two two-level subsystems and found that any pair of atoms in the
ensemble can be inseparable, which gives reason to interpret the
state of the entire system as inseparable.

Note also that spurious entanglement may be predicted by
perturbation theory \cite{28}. In that study, an example of
expansion of the product of two wave functions in terms of a common
classical parameter was considered in which individual summands
represent entangled states. However, if entanglement entropy is used
as a measure, then we have initially independent systems, because
the entropy is either quadratic in the small parameter or zero in
arbitrary-order perturbation theory. Note that physical
implementation of such entangled states, i.e., preparation of an
independent state of a pair of entangled particles, requires
projective measurement in an entangled basis. The present analysis
also relies on perturbation theory, but we deal with a different
situation in both physical and formal sense, in which interaction
between particles gives rise to correlation. The wavefunction
obtained in first-order perturbation theory is not factorizable, and
the corresponding entanglement entropy is zero to the corresponding
accuracy. This result is physically plausible, because there is no
correlation in the first-order perturbation theory. In our analysis,
entanglement is predicted by second-order perturbation theory, which
describes real emission and absorption processes result in
correlation. In this order of perturbation theory, the existence of
quantum correlation is substantiated by entanglement criteria
consistent with approximation accuracy.

\section{Exact solutions}

Radiative decay can be neglected in (\ref{002}) when evolution over
a time $t\ll\gamma^{-1}$ is considered, and the behavior of the
entire system is described by the wavefunction $
\phi(t)=\exp(-i\hbar^{-1}H t)(\phi_{A}\otimes\phi_{f})$, where the
initial states of the atoms and field are assumed to be
uncorrelated. Then, simple solutions can be obtained under certain
initial conditions.

Consider the mixing of modes $a$ and $b$ described by
\begin{eqnarray}
\label{00210} H=i\hbar f(a^{\dagger}bS-ab^{\dagger}S^{\dagger}),
\end{eqnarray}
where $S=S_{10}, S^{\dagger}=S_{01}$. If analysis is restricted to
single-photon Fock states of the modes
$\phi_{f}=c\ket{01}_{ab}+e\ket{10}_{ab}$ exact solutions can be
written as
\begin{eqnarray}
\label{0022} \nonumber
\exp\{-i\hbar^{-1}Ht\}(c\ket{01}_{ab}+e\ket{10}_{ab})\otimes\phi_{A}=c
\Big\{ \ket{01}\cos[tf\sqrt{SS^{\dagger}}]+
\ket{10}S^{\dagger}\frac{1}{\sqrt{SS^{\dagger}}}
\sin[tf\sqrt{SS^{\dagger}}] \Big\}\otimes\phi_{A}&&\\
+e\Big\{-\ket{01}S\frac{1}{\sqrt{ S^{\dagger} S}}
\sin[tf\sqrt{S^{\dagger} S}]+\ket{10}\cos[tf\sqrt{S^{\dagger}
S}]\Big\}\otimes\phi_{A}.&&
\end{eqnarray}
In the case of a single-photon process described by the Hamiltonian
\begin{eqnarray}
\label{00222}
 H=i\hbar g(aS-a^{\dagger}S^{\dagger})
\end{eqnarray}
there also exist simple solutions. For example,
\begin{eqnarray}
\label{0023} \nonumber
\exp\{-i\hbar^{-1}Ht\}(c\ket{1}\otimes\ket{0;N}+e\ket{0}\otimes
\ket{1;N})&&\\
 = c\Big\{\cos[gf\sqrt{N}]\ket{1}\otimes\ket{0;N}
+\frac{1}{\sqrt{N}}\sin[gf\sqrt{N}]\ket{0}\otimes
\ket{1;N}\Big\}&&\\ \nonumber + e\Big\{
-\sqrt{N}\sin[gf\sqrt{N}]\ket{1}\otimes\ket{0;N}+
\cos[gf\sqrt{N}]\ket{0}\otimes \ket{1;N}\Big\}, &&
\end{eqnarray}
where $\ket{h;N}=\ket{0}^{\otimes N}, h=0,1$ represents the ground
state of the atomic ensemble and a symmetric Dicke state defined in
accordance with (\ref{0021}). These solutions are valid only under
the restrictions imposed above on the initial states. They describe
exchange of excitation between the cavity mode and the atoms.

\section{Generation and transformation of symmetric states }

Now, we use the exact solutions written out above to analyze the
evolution of symmetric Dicke states  $\ket{h;N}$ in single- photon
and wave-mixing processes.

First, consider the case when the spatial inhomogeneity of the field
within the region occupied by the atomic ensemble can be neglected.
Setting (\ref{0023}) $\phi_{A}=\ket{h;N}$, we use (\ref{DD}) to
obtain
\begin{eqnarray}
\label{005}
 \nonumber
\Big(\alpha\ket{01}+\beta\ket{10}\Big)\otimes\ket{h;N}\to%&&\\
\nonumber
\alpha\Big\{\cos\theta_{h}\ket{01}\otimes\ket{h;N}+\sqrt{\frac{h+1}{N-h}}
\sin\theta_{h}\ket{10}\otimes\ket{h+1;N}\Big\}&&\\
+\beta\Big\{-\sqrt{\frac{N-h+1}{h}}\sin\theta'_{h}\ket{01}\otimes
\ket{h-1;N}+\cos\theta'_{h}\ket{10}\otimes\ket{h;N}\Big\}&&
\end{eqnarray}
where $\theta_{h}=tf\sqrt{(h+1)(N-h)}$,
$\theta'_{h}=tf\sqrt{h(N-h+1)}$. Relation (\ref{005}) entails
possibilities of preparation of an entangled from ground-state atoms
$ \ket{0;N}\to\ket{1;N},$ and transformation of entangled states by
changing the number of excited atoms $ \ket{h;N}\to\ket{h\pm1;N},$
including disentanglement: $ \ket{h;N}\to\ket{h-1;N}\to\dots
\ket{0;N}.$

Note that exact solutions (\ref{0023}) and (\ref{005}) describe
state swapping, which can be used to map the state of light onto
atoms in order to store it in a long-lived atomic ensemble, i.e., to
implement quantum memory. In particular, an unknown superposition of
photons can be transferred to atoms and back by using the following
transformation entailed by (\ref{0023})
\begin{equation}\label{5210}
\Big(\alpha\ket{1}+\beta\ket{0}\Big)\otimes\ket{0;N}\leftrightarrows
\ket{0}\otimes\Big(\alpha\frac{1}{\sqrt{N}}\ket{1;N}+\beta\ket{0;N}\Big).
\end{equation}
Solutions (\ref{0023}) and (\ref{005}) make it possible to take into
account the spatial configuration of atoms in the ensemble. For
example, consider the interaction between a one-dimensional array of
atoms located at points $x_{1},\dots,x_{N}$ and a single photon
described by Hamiltonian (\ref{00222}) with
$S=\sum_{p}s_{10}(p)\exp[ikx_{p}]$, where
$s_{10}(p)=\ket{1}_{p}\bra{0}$ corresponds to the atom located at
$x_{p}$, $p=1,\dots,N$. Using (\ref{0023}), we can show that
\begin{eqnarray}
  \ket{1}\otimes\ket{0;N}\to\cos\theta \ket{1}\otimes\ket{0;N}+
  \sin\theta\ket{0}\otimes\eta_{N}(1),
\end{eqnarray}
где $\theta=tg\sqrt{N}$,
\begin{eqnarray}
\label{0055}
\eta_{N}=(1/\sqrt{N})\Big[e^{ikx_{1}}\ket{10\dots0}+\dots
e^{ikx_{N}}\ket{0\dots 01}\Big].
\end{eqnarray}
Expression (\ref{0055}) implies that an array of entangled atoms is
created when $\theta=\pi/2$. Note that $\eta_{N}$ is the Dicke state
with $j=m=N/2-1$ only if $\sum_{p}\exp[ikx_{p}]=0$.

\section{Entangled atomic ensembles}

Solutions (\ref{0022}), (\ref{0023})) imply that a photon and an
atomic ensemble are entangled via interaction. If photons are
entangled (e.g., by projective measurement) in a combination of such
independent systems, then the atomic ensembles will become
entangled. We consider optical measurement schemes based on this
method, known as entanglement swapping. The key resources used in
these schemes are set of atomic ensembles correlated with respective
photons, beamsplitters, and single-photon detectors. The analysis
that follows is restricted to schemes in which only specific
single-photon output is recorded.

As an initial state, we use the EPR pair
\begin{eqnarray}
\label{004}
Z(W)=a\ket{0}_{f}\otimes\ket{0}+b\ket{1}_{f}\otimes\ket{W},
\end{eqnarray}
where Fock states are denoted by the subscript “f,”
$\ket{W}=\ket{1;N}/\sqrt{N}, \ket{0}=\ket{0;N}$. It is generated by
the mode mixing described by (\ref{00210}), where the mode $b$ is a
classical wave. The state of $n$ independent identical ensembles
entangled with respective photons is represented by the product
\begin{eqnarray} %\nonumber
\label{0041} Z_{n}(W)=Z(W)^{\otimes
n}%&&\\
=a^{n-1}b\Big[\ket{10\dots0}_{f}\otimes\ket{W0\dots 0}+\dots
\ket{00\dots 1}_{f}\otimes\ket{00\dots W}\Big]+\dots&&
\end{eqnarray}
As illustrated by the figure, the photons associated with atomic
ensembles are injected into a system of $n-1$ beamsplitters with $n$
input ports and n output ports.
\begin{figure}[h]
\label{Bs}
  \centering
\epsfxsize=9cm \epsfbox{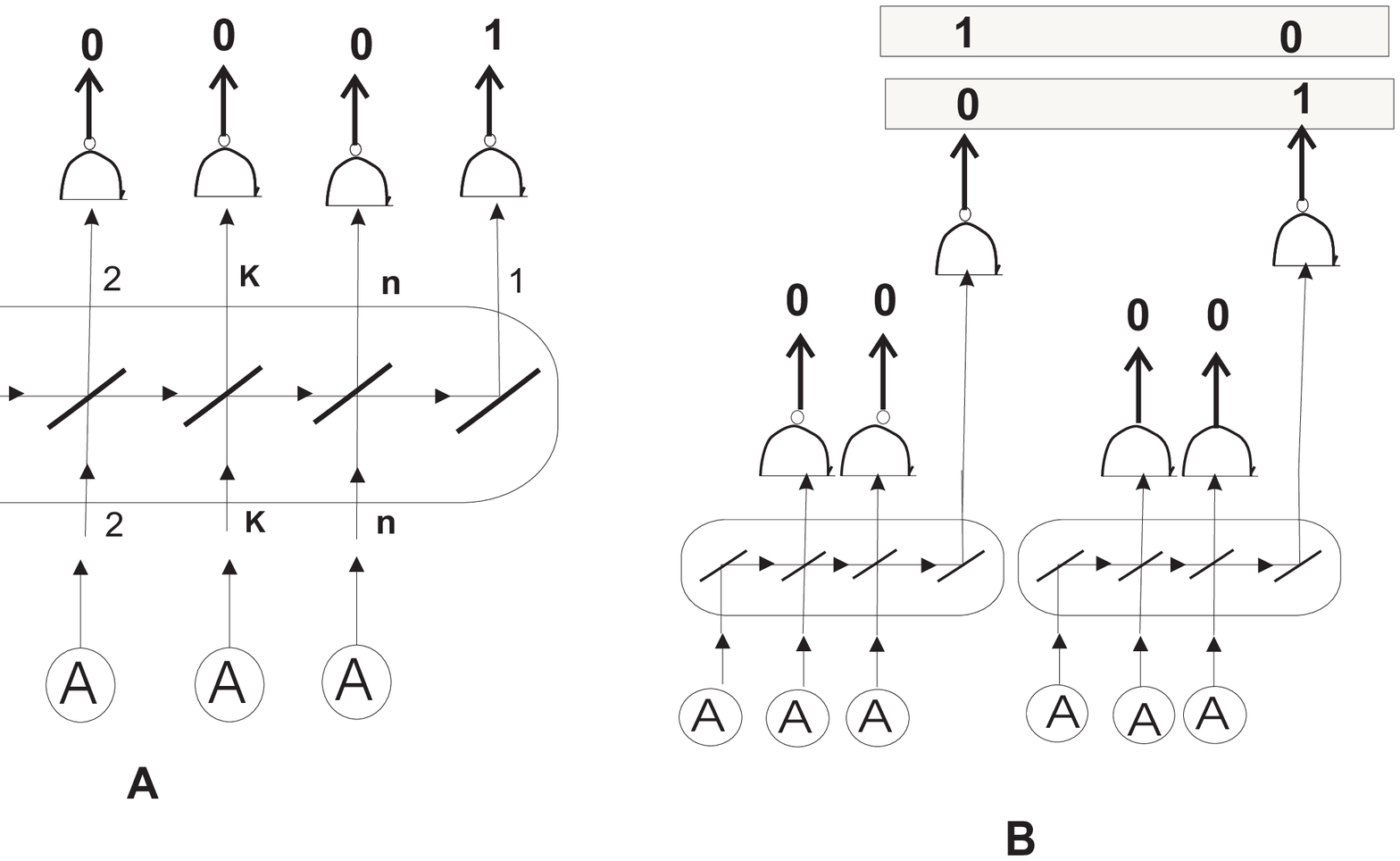}
  \caption{(a) Scheme for generating entangled states of atomic
ensembles. (b) Preparation of entangled states by correlation of
photocounts recorded by two schemes.}
\end{figure}
Each beamsplitter performs the transformation $\ket{01}_{f}\to
c_{k}\ket{01}_{f}+s_{k}\ket{10}_{f}$,
$\ket{10}_{f}\to-s_{k}\ket{01}_{f}+c_{k}\ket{10}_{f}$, where
$c_{k}^{2}+s_{k}^{2}=1$, $k=1\dots n-1$. The scheme is described by
a unitary operator $U_{nf}$ and characterized by the following
property. There exist an input port optically coupled to every
output port and an output port optically coupled to every input
port. In Fig. 1a, the latter is output port 1. We call it the
optical output port, and the corresponding detector is called the
output detector. The scheme performs the transformation
\begin{eqnarray}
\nonumber
 U_{nf}\ket{1\dots 0}_{f}=t_{1}\ket{1\dots 0}_{f}+\dots
+t_{n}\ket{0\dots 1}_{f},&&\\
\label{0042}
 U_{nf}^{-1}\ket{1\dots 0}_{f}=\tau_{1}\ket{1\dots
0}_{f}+\dots +\tau_{n}\ket{0\dots 1}_{f},&&
\end{eqnarray}
where the coefficients $t_{k},\tau_{k}$, $k=1,\dots n$ are
determined by the transmittances and reflectances of the
beamsplitters, and $\sum_{k}t^{2}_{k}= \sum_{k}\tau^{2}_{k}=1$.

If output detector detects a photon which corresponds to the state
$\ket{1_{f}}=\ket{1\dots 0}$, then with the probability
\begin{eqnarray}
\label{000ver}
 Prob(1)=|a^{n-1}b|^{2}
\end{eqnarray}
entangled state of atomic ensembles will be prepared
\begin{eqnarray}
\nonumber
 \label{0043}
\bra{1_{f}}U_{nf}Z_{n}(W)/\sqrt{Prob(1)}=\eta_{n}(W),&&\\
\eta_{n}(W)=q_{1}\ket{W\dots 0}+q_{n}\ket{0\dots W},&&
\end{eqnarray}
This scheme has the following property. Since the coefficients
$q_{1},..q_{n}$  are completely determined by the transmittances and
reflectances of the beamsplitters, weakly entangled states $Z(W)$
can be used to prepare highly entangled states atomic ensembles.

Let us consider several particular cases. If $n = 2$, then
$q_{1}=c_{1},q_{2}=s_{1}$, and we have an EPR pair of the form $
\eta_{2}(W)=EPR(W)=c_{1}\ket{W0}+s_{1}\ket{0W}$. When
$c_{1}=s_{1}=1/\sqrt{2}$ it is maximally entangled. If $n = 3$ and
$q_{1}=c_{1}c_{2},q_{2}=-s_{1}c_{2},q_{3}=s_{2}$, then we have a $W$
state. If $c_{1}=-s_{1}=1/\sqrt{2}$, $c_{2}=\sqrt{2/3}$ and
$c_{2}=\sqrt{2/3}$, then
\begin{eqnarray}
\label{0047}
\eta_{3}(W)=W(W)=(1/\sqrt{3})(\ket{W00}+\ket{0W0}+\ket{00W}).
\end{eqnarray}
In particular, one can prepare the asymmetric state
$\widetilde{W}(W)=(1/\sqrt{2})\ket{W00}+(1/2)\ket{0W0}+(1/2)\ket{00W}$.
When $N = 1$, it is unitary equivalent to the GHZ state and can be
used as a quantum channel for teleportation or dense coding
\cite{29}.

Using correlation between photocounts in a combination of schemes
considered above, mixed states of atomic ensembles can be prepared,
including inseparable ones. For example, consider two independent
identical schemes $S_{2}(X)$ combined as shown in Fig. 1b, with
three single-photon detectors in each scheme. If a photon is
detected by either scheme, then we have the pair of states $
\bra{1_{f}}S_{2}(X)\otimes\bra{0_{f}}S_{2}(X)w=\ket{\eta_{2}(X),0}$
and $
\bra{0_{f}}S_{2}(X)\otimes\bra{1_{f}}S_{2}(X)w=\ket{0,\eta_{2}(X)}
$. Suppose that the detector outputs are connected so that a single
photon produced by either scheme is counted. This measurement is
described by the projector $\proj{1_{f}0_{f}}+\proj{0_{f}1_{f}}$.
The resulting mixed state is represented by a density matrix of the
form
\begin{eqnarray}
\label{0057}
\rho(X)=(1/2)\Big[\proj{\eta_{2}(X),0}+\proj{0,\eta_{2}(X)}\Big].
\end{eqnarray}
Its separability is an open question, because a necessary and
sufficient condition is known only for mixed systems of dimension
$2\times 2, 2\times 3$. However, if we assume that $N = 1$, i.e.,
consider a combination of four atoms instead of ensembles, then
$\eta_{2}(X)=\Psi^{+}=(1/\sqrt{2})(\ket{01}+\ket{10})$ and density
matrix (\ref{0057}) describes a four-particle state:
\begin{eqnarray}
\rho(4)=(1/2)(\proj{\Psi^{+}00}+\proj{00\Psi^{+}}).
\end{eqnarray}
Taking the state of the pair of atoms in the first scheme defined by
the two-particle reduced density matrix
$\rho(2)=(1/2)\Big[\proj{\Psi^{+}}+\proj{00}\Big]$, we can apply the
separability criterion. The density matrix partially transposed over
the variables of the one atom has four eigenvalues one of which is
negative $1/4,1/4, (1\pm\sqrt{2})/4$. Therefore, the density matrix
$\rho(4)$ is inseparable.

\section{Hierarchic structure of states}

Note that expression (\ref{0043}) is hierarchically structured. To
illustrate this property, we consider a combination of schemes
generating states of this type. As distinct to schemes using
correlation of photocounts, we consider optically connected schemes.
If an elementary scheme that performs the transformation $S_{n}(X) =
U_{nf}Z_{n}(W)$ with $X = W$ (see Fig. 1a) records single-photon
output, then the resulting state has the form of (\ref{0043}):
\begin{eqnarray}
\label{591}
 \bra{1_{f}}S_{n}(X)w=\eta_{n}(X)=\tau^{`}_{1}\ket{X0\dots
0}+\dots+\tau^{`}_{n}\ket{00\dots X},
\end{eqnarray}
where $w=1/\sqrt{Prob(1)}$. We define the optical output port of the
scheme $S_{n}(X)$ as the one optically coupled to every input port.
In Fig. 1a, it is output port 1. The input port of the scheme
$S_{n}(X)$ is defined as the optical input port of the system of
beamsplitters. Then, we can take, for example, p independent schemes
represented as $(S_{n}(X))^{p}$ and use their optical outputs as the
input of the scheme $S_{p}$. As a result, we have a new scheme
$S_{p}((S_{n}(X))^{p})$. If it records single-photon output, we have
an entangled state that consists of lower level entangled states:
\begin{eqnarray}
\bra{1_{f}}S_{p}((S_{n}(X))^{p})w=\eta_{p}(\eta_{n}(X))
=t_{1}\ket{\eta_{n}(X),0\dots
0}+\dots+t_{p}\ket{0,0,\dots\eta_{n}(X)}.
\end{eqnarray}
By virtue (\ref{591}) it takes the forme:
\begin{eqnarray}
\eta_{p}(\eta_{n}(X)) =\eta_{pn}(X).
\end{eqnarray}
\\
Thus, we can formulate the following property. The state
$\eta_{n}(X)$ defined by (\ref{0047}) with $n=n_{1}n_{2}\dots n_{p}$
can be represented as
\begin{eqnarray}
\eta_{n}(X)=\eta_{n_{1}}(\eta_{n_{2}}(\dots(\eta_{n_{p}})).
\end{eqnarray}
This implies that the vector $\eta_{n}(X)$ has the structure of an
entangled state with respect to any group of $s$ particles, where s
is such that $n/s$ is a natural number greater than unity.

When the wavefunction $\eta_{n}(X)$ is symmetric, a hierarchically
structured representation can be obtained by using the permanent
expansion defined as a determinant with a summation rule for
permutations depending on symmetry \cite{30}. In particular,
successive decomposition of a determinant with respect to rows or
columns and subsequent association of summands can be used to
represent a permanent in terms of permanents of lower dimension,
which reflects hierarchical structure.

For example, when $n = 6$, it holds that
\begin{eqnarray}
\eta_{6}(X)=\eta_{3}(\eta_{2}(X))=\eta_{2}(\eta_{3}(X))).
\end{eqnarray}
This state has the structure of an EPR pair or a W state:
\begin{eqnarray}
\nonumber \eta_{3}(\eta_{2}(X))=W(EPR)=EPR(W).&&
\end{eqnarray}
This example demonstrates that the same state exhibits structure
characteristic of entangled states of two different types. This
property can be used in different applications: the EPR pair can
serve as a quantum channel for teleportation or dense coding, while
the symmetric W state can be used for cloning.

To choose a particular structure defined by the dimension of the
Hilbert space of its element, appropriate basis vectors and
observables should be introduced. In physical terms, this is
equivalent to a two-level approximation. Indeed, any group of $s$
particles, where $s$ is such that $n/s$ is a natural number greater
than unity, is represented in $\eta_{n}(X)$ by two states,
$\ket{0}=0_{s}$ и $\eta_{s}(X)=1_{s}$. The group can be treated as a
two-level particle (qubit) with basis vectors $0_{s}$ and $1_{s}$.
Such qubits and hierarchically structured states $\eta_{n}(X)$ can
be used in quantum information processing. By analogy with
(\ref{0055}), the vector $\eta_{n}(X)$ represents a Dicke state only
if $\eta_{n}(X)$.

\section{Conclusions}

A model describing resonant interaction of identical two-level atoms
with a narrow-band radiation field is used to analyze multiparticle
entanglement. The interaction is described by an effective
Hamiltonian that allows for various multiphoton processes. The
statistics of radiation and atoms are characterized by a density
matrix, for which solutions are calculated in secondorder
perturbation theory in the interaction strength and exact solutions
are found in the case of negligible decay. It is shown that the
state of any pair of atoms interacting with weak single-mode
squeezed light is inseparable and robust against decay. It is
demonstrated that symmetric entangled multiparticle states can be
generated by using optical schemes based on singlephoton projection.
An optical scheme is described that can be used to prepare highly
states of entangled atomic ensembles from weakly entangled states by
projective measurement.

This work was supported, in part, by Delzell Foundation, Inc.

\end{document}